\documentclass[12pt,a4paper]{panl}
\usepackage{graphicx}
\usepackage{amssymb}
\usepackage{amsfonts}
\usepackage{amsmath}
\usepackage{longtable}
\usepackage{lscape}
\usepackage{epsfig}
\usepackage{multirow}

\originalTeX
\begin{document}

\title{ Dependence of the structure  of the  elastic scattering amplitude on Mandelstam variables at high energies}
\maketitle
\authors{O.V.\ Selyugin$^{a}$\footnote{E-mail: selugin@theor.jinr.ru}
}
\setcounter{footnote}{0}
\from{$^{a}$\,BLTP JINR, Dubna, Russia}

\begin{abstract}
  Analysis of new experimental data obtained by the TOTEM  and ATLAS Collaborations at the LHC
 together with old data obtained at the SPS and Tevatron colliders  at small momentum transfer
in the framework of the high energy generalized structure (HEGS) model allows one to determine
 the dependence of different parts   of the hadron elastic scattering amplitude
  on the mandelstam kinematic variables the $s$ and $t$  .

\end{abstract}
\vspace*{6pt}


\label{sec:intro}
\section{Introduction}
There are   some problems   as confinement, hadron interaction at large distances, non-perturbative hadron structure   parton distribution functions (PDFs),
 generalized parton distributions (GPDs) and others
 that    should be explored in the framework of the Standard Model. These problems are determined, in most part,  by the non-perturbative hadron interactions at large distances
  and  connected with the hadron interaction at high and super-high energies   and with the problem of  energy dependence of the structure
 of the  scattering amplitude and  total cross sections   \cite{royt,Rev-LHC,Block-85}.
  This reflects   a tight connection of the main properties of  elastic hadron scattering   with the first principles of quantum field theory  and the concept of the scattering amplitude as a unified analytic function of its kinematic variables
which were introduced by   N.N. Bogolyubov \cite{Bogolyubov:1983gp}.
Researches into the structure of the elastic hadron scattering amplitude   at super-high energies and small momentum transfer $t$    outline the connection   between    experimental knowledge and  fundamental   asymptotic theorems that are based on first principles.
  The dispersion relation tightly connects the real and imaginary parts of the elastic scattering amplitude. Hence  the $t$ and $s$ dependence of the real part is determined by the imaginary part of the hadron elastic scattering amplitude.

 This connection provides    information about  hadron interaction
  at large distances    where the perturbative QCD does not work,
   which gives the premise for a new theory to be developed.
  The new experimental data on elastic proton-proton scattering  obtained at the LHC by the TOTEM (SMS) \cite{T7a,T8,8T,T13-1set,T13-2set,T2p76}
  and the ATLAS Collaborations \cite{AT7,AT-14,AT78,AT16,AT17,AT19,AT13} essentially expanded our knowledge about hadron interaction at super-high energies.

\label{sec:experiment}
\section{The elastic hadron scattering amplitude in the framework of the High energy generalized structure  (HEGS) model and the differential cross section at high energies }

\begin{figure}[t]
\begin{flushright}
\includegraphics[width=127mm]{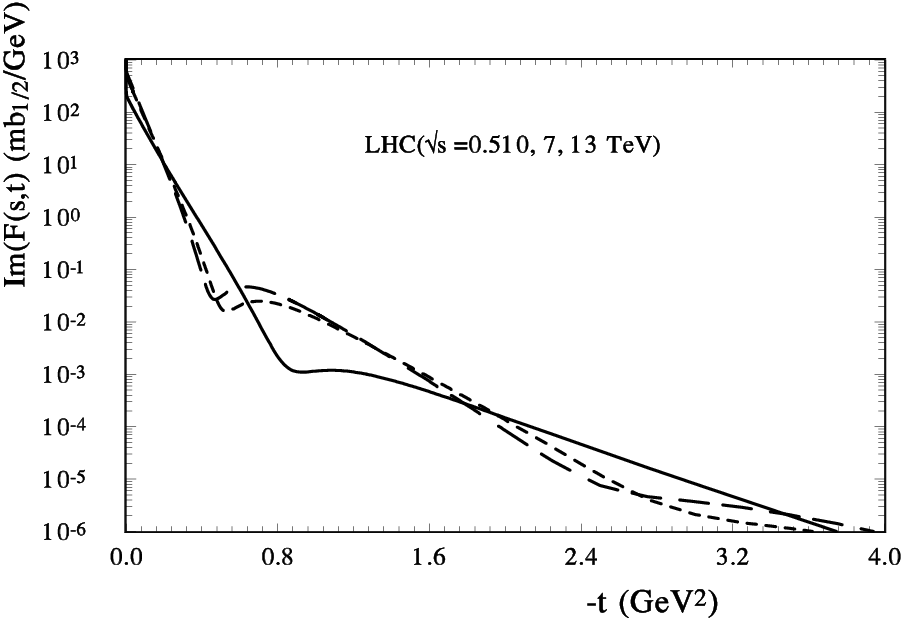}
\vspace{25mm}
\caption{The differential cross sections of elastic proton-proton scattering at
  $\sqrt{s} =  0.51, 7, 13$ TeV are calculated in the model (points - the experimental data) }
\end{flushright}
\labelf{fig01}
\end{figure}

  Now our model describes experimental data on elastic differential cross sections
  in the maximum wide range of energy, from low energy $\sqrt{s}=6 $ GeV up to the recent
   maximal energy $\sqrt{s}=13 $ TeV
  obtained at accelerators \cite{data-Sp,Land-Bron}
  simultaneously and in a wide momentum transfer (including the Coulomb nuclear interference region, diffraction dip-bump structure and
  the large momentum transfer (up to $|t|=15$ GeV)  \cite{HEGS0,HEGS1}.
  The model takes into account the pomeron (cross-even) and odderon (cross-odd) parts of exchange
    in the $t$-channel. Of course, it also takes into account all five helicity electromagnetic amplitudes and the Coulomb hadron interference term. This allows us to obtain a quantitative description
    of the differential cross section at super small momentum transfer \cite{HEGSh,HEGS-AT13}.

  In Fig. 1,  the differential cross sections of elastic proton-proton scattering at
  $\sqrt{s} =  0.51, 7, 13$ TeV  calculated in the model are present.
   It can be seen that at low momentum transfer there exists a crossover point of momentum transfer.
    It reflects the growth of the total cross section  on one hand and the growth of the slope
   of the differential cross section on the other hand.


\begin{figure}[t]
\begin{flushright}
\includegraphics[width=127mm]{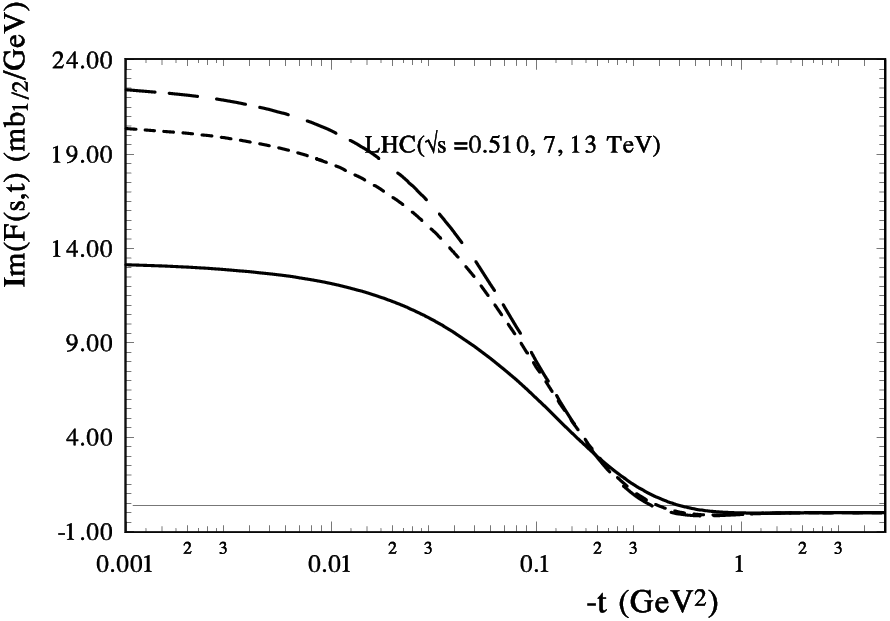}
\vspace{25mm}
\caption{The dependence of the imaginary part of the hadron scattering amplitude on $s$ and $t$
  calculated in the model at the energy  $\sqrt{s} =  0.51, 7, 13$ TeV }
\end{flushright}
\labelf{fig01}
\end{figure}

    The position of the diffraction minimum $t_{min}(s,t)$ moves to low momentum transfer continuously  \cite{HEGS-min}.
   It is interesting that the velocity of changing the position of the diffraction minimum
   changes very slowly. For example, from ISR energy $\sqrt{s}=53 $ GeV
   up to SPS energy $\sqrt{s}=540 $ GeV such position changes with a speed of $0.11$ GeV$^2$
    per $100$ GeV. Between $\sqrt{s}=540 $ GeV and $\sqrt{s}= 7 $ TeV
    such speed  is two times less and equals $0.006 $, at last between $7$ and $13$ TeV the position of minimum changes with a speed
    of  $0.002$ per $100$ GeV.  Approximately, scaling of this process can be represented as
       $t_{min} \ln{s/s_{0} } = const$.
    After the second bump the slope of the differential cross sections increases with energy.
    It  corresponds to the grows of the slope of the diffraction peak.

\begin{figure}[t]
\begin{flushright}
\includegraphics[width=127mm]{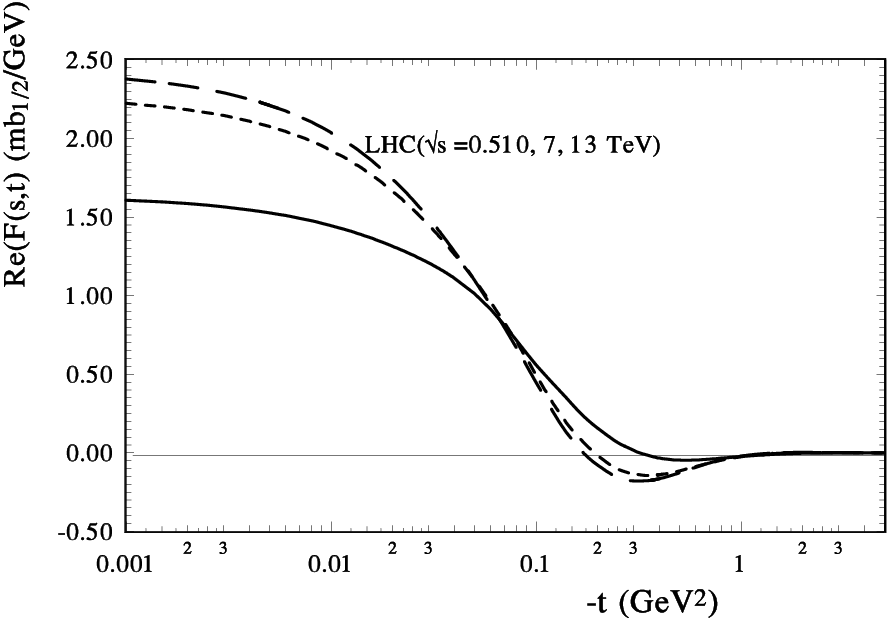}
\vspace{25mm}
\caption{The dependence of the real part of the hadron scattering amplitude on $s$ and $t$
  calculated in the model  at the energy  $\sqrt{s} =  0.51, 7, 13$ TeV }
\end{flushright}
\labelf{fig01}
\end{figure}

 The behavior of the imaginary part of the scattering amplitude over momentum transfer
    is presented in Fig.2 for the energies $\sqrt{s} =  0.51, 7, 13$ TeV.
    Again, we can see the point of crossover in the region of $|t|=0.2$ GeV$^2$.
    Despite the essential grows of the size of the imaginary part of the scattering amplitude at very small momentum transfer, its slope slightly  changes with $t$ in the region of the  Coulomb nuclear interference. The size of slope is practically proportional to the size of the total cross section in that region. However at larger $t$, for example at $|t|=0.1$ GeV$^2$,
    it grows essentially faster.

    It should be noted that the size of the slope of the differential cross sections  is determined in that region of $t$ by the CNI interference term which is proportional
    to $\alpha/t$  . It allows us to analyze  \cite{HEGS-AT13} the first points of the unique experiment carried out by the ATLAS Collaboration \cite{AT13}.
     The point of $t$, where the imaginary part changes its sign, determines the position of the diffraction minimum. But it slightly moves at some large $t$ by the contribution of the real part of the elastic hadron scattering amplitude.

\begin{figure}[t]
\begin{flushright}
\includegraphics[width=127mm]{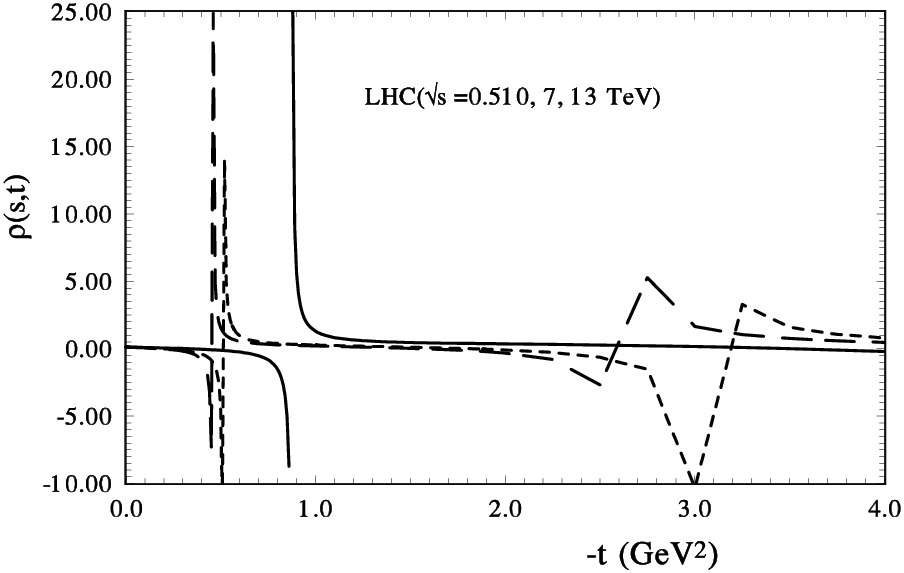}
\vspace{25mm}
\caption{The size of the $\rho(s,t)$ - ratio of the real to imaginary part of the hadron scattering amplitude is calculated in the model at the energy  $\sqrt{s} =  0.51, 7, 13$ TeV depending on $s$ and $t$.  }
\end{flushright}
\labelf{fig01}
\end{figure}

  It is interesting that the form of the real part of the hadron elastic scattering amplitude
   it is similar to its imaginary part. Of course, they are not proportional to each other as their connection has to satisfy the dispersion relations \cite{CS-PRL} which require, for example, the changing size of the real part at sufficiently small momentum transfer. Really, in Fig.3, we see that the real part change its sign   essentially earlier than the position of the diffraction minimum.
   At LHC energies this happens in the area of momentum transfer approximately equal to $0.2$ GeV$^2$. But before that, we can see the crossing point at $|t| \sim 0.06$ GeV$^2$.
   Likely, the behavior of the imaginary part the slope of the real part
   at very small momentum transfer is also practically proportional to the size of the total cross section at different energies and grows at larger momentum transfers. It can be see that the real part
    at LHC energies has the negative maximum at approximately $|t|=0.3$ GeV$^2$ situated near the diffraction minimum. Hence, it  essentially impacts  the form and size of the diffraction minimum in the differential cross sections.  In Fig. 4, the ratio $\rho(s,t)$ of the real to imaginary part  of
    the hadron elastic scattering amplitude is presented for different energies.
    Such a complicated structure of   $\rho(s,t)$ are determined by the changes of the sign of the real and
    imaginary parts of the scattering amplitude. At small momentum transfer, the size of $\rho(s,t)$ is small
    as the real part changes its sign. Contrary, when the imaginary part changes its sign, the size of $\rho(s,t)$ grows very faster. The energy dependence of $\rho(s,t)$ is due to the movement of the position of the diffraction minimum, hence with the energy dependence of the imaginary part of the scattering amplitude.

  In the HEGS model, the parameters are determined by the minimization of $\chi^2$ \cite{Sitnik1,Sitnik2} with fitting simultaneously
  all the existing experimental data. It should be noted that in such a fitting procedure the model used only
  statistical errors of the experimental data. The systematic errors are used as  additional normalization which is equal to all points of one experimental set. It essentially decreases the space for the fitting
  theoretical curve. This method allows one to discover two new effects in the differential cross section:
   oscillations of the differential cross section  at small angles  \cite{osc-13},
   which analyzed in \cite{Per-1}   and an additional term with
 a large slope in the scattering amplitude \cite{fd13,fd-LHC}.
As a result, the model calculations give  at $\sqrt{s} = 7 $ TeV the size of the total cross section
equal to $99.5 \pm 1.5$ mb and the size of the ratio real part to imaginary part of the elastic hadron scattering amplitude $\rho(t=0)=0.106 \pm 0.025$.  At $\sqrt{s} = 13$ TeV, the model gives
 the same values $109.8 \pm 1.46$ mb and $\rho(t=0)=0.111 \pm 0.024$.

\section{Conclusion}

In our HEGS model the hadron elastic scattering amplitude  has the analytic properties in the framework  of Mandelstam variable $s$ and $t$  and  satisfies the dispersion relations.
As a result, the real part of the scattering amplitude is determined by the complex
$\hat{s} = s e^{i\pi/2}$. It is very important that
the basic terms of the scattering amplitude (Pomeron and Odderon)
   has the same intercept. The model used two forms of form factors.
   The pomeron amplitude is proportional to the electromagnetic form factor, but the odderon
   amplitude is proportional to the gravitation form factor. Both form factors were calculated from the
   same generalized parton distributions $GPDs(x,t)$.
  The extended variant of the model shows the contribution of the “maximal” odderon with specific kinematic properties.
The final scattering amplitude, obtained by eikonalization,
   satisfies the unitarity and Froiasart-Martin limit.
   The model satisfies  all analytic properties required by quantum field theory.
In framework of the model, new effects in
   the elastic hadron scattering at small momentum transfer
   (oscillations and some part with large slope) was determined.
The model opens a new way to determine the true form of the GPDs and standard parton distributions.

\vspace{1cm}

  FUNDING \\
This research was carried out at the expense of the grant of the Russian Science Foundation No. 23-22-00123, https://rscf.ru/project/23-22-00123 /. \\

\vspace{0.5cm}

CONFLICT OF INTEREST \\
The authors declare that they have no conflicts of interest.\\


\end{document}